\documentclass[12pt,aps]{revtex4}
\usepackage{amsfonts,amsbsy,bm}
\usepackage{graphicx} 

\begin{document}

\title{Quantum brachistochrone problem for spin-1 in a magnetic field}

\author{A.~M.~Frydryszak$^1$, V.~M.~Tkachuk$^2$}
\address{$^1$Institute of Theoretical Physics, University of
Wroclaw,
 pl.~M.~Borna 9, 50--204 Wroclaw, Poland \\
 e-mail: amfry@ift.uni.wroc.pl \\
 $^2$Ivan Franko Lviv National University,
 Chair of Theoretical Physics
  12 Drahomanov Str., Lviv UA--79005, Ukraine \\
    e-mail: tkachuk@ktf.franko.lviv.ua}

\begin{abstract}
We study quantum brachistochrone problem for the spin-1 system in a
magnetic field of a constant absolute value. Such system gives us a
possibility to examine in detail the statement of papers [A.
Carlini {\it et al.}, Phys. Rev. Lett. {\bf 96}, 060503 (2006), D.
C. Brody, D. W. Hook, J. Phys. A {\bf 39}, L167, (2006)] that {\it
the state vectors realizing the evolution with the minimal time of passage
evolve along the subspace spanned by the initial and final state vectors.}
Using explicit example we show the existence of quantum
brachistochrone with minimal possible time, but the state vector of
which, during the evolution {\em leaves} the subspace spanned by the initial and final state vectors. This is the result of the choice of more constrained Hamiltonian then assumed in the general quantum brachistochrone problem, but what is worth noting, despite that such evolution is more complicated it is still time optimal. This might be important for experiment, where general Hamiltonian
with the all allowed parameters is difficult to implement, but constrained one depending on magnetic field can be realized. However for pre-constrained Hamiltonian not all final states are accessible.
Present result does not contradict general statement of the quantum brachistochrone problem, but gives new insight how time optimal passage can be realized.

\end{abstract}

\maketitle

Recently Carlini {\it et al.} \cite{Carlini06} generalized the
classical brachistochrone problem to the quantum case. The
quantum brachistochrone problem can be formulated in the following way: What is the
optimal Hamiltonian, under a given set of constraints such that
the evolution from a given initial state $|\psi_i\rangle$ to given
final one $|\psi_f\rangle$ is achieved in the shortest time. Using
the variational method the authors solved this problem for some
specific examples of constrains which lead to fixed distance
between the largest and the smallest energy levels of the
Hamiltonian. In \cite{Brody167} it was shown that quantum
brachistochrone problem can be solved more directly using the
symmetry properties of the quantum state space. That paper was
based on the idea considered in \cite{Brody5587}, where an
elementary derivation was provided for passage time from the
one quantum state into another orthogonal one.

Later the variational method was extended to allow finding the
time-optimal realization of a target unitary operation, when the
available Hamiltonians are subjected to certain constrains
dictated either by experimental or theoretical conditions
\cite{Carlini04}. In \cite{CarliniPr47} the authors considered the
brachistochrone problem for quantum evolution of mixed states.
Very recently Bender {\it et al.} studied the brachistochrone
problem for a PT-symmetric non-Hermitian two-dimensional matrix
Hamiltonian \cite{Bender040} and showed that among non-Hermitian
PT-symmetric Hamiltonians satisfying the same energy constraint
the time evolution between two fixed states can be made
arbitrarily small. Such an interesting phenomenon was observed
also for dissipative systems described by a non-Hermitian
Hamiltonian which has a negative imaginary part of eigenvalues
\cite{AssisPr54}. Discussion on this subject can be found also in
\cite{Martin23,Mosta44}.

Important statement of work \cite{Carlini06,Brody167} (see also
\cite{Bender040}) is that {\it finding the minimal time of
general evolution reduces to finding optimal time evolution for the
Hamiltonian acting on the two-dimensional subspace spanned by the
initial and final state vectors $|\psi_i\rangle$ and
$|\psi_f\rangle$}. It means that optimal evolution which realizes
quantum brachistochrone can be written as a linear combination of
$|\psi_i\rangle$ and $|\psi_f\rangle$ with time dependent
coefficients. One of the aims of our paper is to examine this
statement in detail within the three-dimensional quantum
system.

We consider the brachistochrone problem in the case when optimal
Hamiltonian belongs to the pre-constrained class of Hamiltonians,
less general, with a less then allowed number of free parameters
that can be used for the problem. Such case is important from the physical
point of view when an experimentalist has a possibility to change
only a few parameters of Hamiltonian but not all. As an example of
such a scenario we consider a three level system, namely, spin-1 in
the external magnetic field described by the Hamiltonian of the
following form
\begin{eqnarray} \label{H}
H=\hbar\omega ({\bf n}\cdot{\bf s}),
\end{eqnarray}
where $\bf s$ are dimensionless spin-1 operators, $\bf n$ is the
direction of the magnetic field and $\hbar\omega$ is proportional
to the strength of the magnetic field. Eigenvalues of this
Hamiltonian are $-\hbar\omega$,  $\hbar\omega$, and $0$. The
difference between the largest and the smallest eigenvalues is
fixed $\Delta E=\hbar\Delta\omega=2\hbar\omega$, what corresponds
to fixed absolute value of magnetic field.

The Hamiltonian (\ref{H}) contains only two free parameters,
namely, two angles $\theta$ and $\phi$ which set the direction of
the magnetic field
\begin{eqnarray}
n_x=\sin\theta\cos\phi,\ \ n_y=\sin\theta\sin\phi,\ \
n_z=\cos\theta.
\end{eqnarray}
Note that general Hamiltonian in three-dimensional space
can be represented by a $3\times 3$ Hermitian matrix which
contains nine free parameters (eight if we consider $su(3)$ case).
We consider here the pre-constrained class of Hamiltonians (\ref{H})
with only two free parameters.

The brachistochrone problem in this restricted case reads: What is the
optimum choice of the pre-constrained Hamiltonian, namely, what is the optimal direction of the
magnetic field $\bf n$ at fixed $\omega$, such that the evolution
from a given initial state $|\psi_i\rangle$ to a given final one
$|\psi_f\rangle$ is achieved in the shortest time?
Obviously, with such a restriction of possible evolutions not all general
final states can be reached. This is the price for taking the narrower family
of Hamiltonians. However,  as we show it below the shortest time achieved in optimal evolution is the same as in the general setting despite the fact that evolution in our case is more complicated (system leaves the subspace spanned on the intial and final states). Let us see it in detail.

The vector of state for spin-1 contains four parameters.
We can write
\begin{eqnarray}\label{abc}
|\psi\rangle=\left(\begin{array}{c}
               a \\
               b \\
               c
             \end{array}\right)=
             \left(\begin{array}{c}
               |a|e^{i\alpha_1} \\
               |b|e^{i\alpha_2} \\
               |c|e^{i\alpha_3}
             \end{array}\right)=e^{i\alpha_1}
             \left(\begin{array}{c}
               |a| \\
               |b|e^{i\alpha} \\
               |c|e^{i\alpha'}
             \end{array}\right),
\end{eqnarray}
where normalization condition is the following: $|a|^2+|b|^2+|c|^2=1$. Hence
four independent parameters, for instance $|a|$, $|b|$, $\alpha$,
and $\alpha'$ define the quantum state. Therefore, having only two parameters in the Hamiltonian to change we
cannot reach arbitrary quantum state starting from a given initial
one, in other words, the evolution defined by the Hamiltonian (\ref{H})
cannot relate two arbitrary quantum states. In our restricted quantum brachistochrone problem we consider only the states which can be
connected by the implemented pre-constrained evolution.

The evolution of the state vector can be realized as follows
\begin{eqnarray} \label{psit}
|\psi(t)\rangle=e^{-iHt/\hbar}|\psi_i\rangle=e^{-i\omega ({\bf n}\cdot{\bf s})t}|\psi_i\rangle.
\end{eqnarray}
It is convenient to represent the unitary operator of evolution in
the form
\begin{eqnarray}\label{Exp}
e^{-i\omega ({\bf n}\cdot{\bf s})t}=
1-({\bf n}\cdot{\bf s})^2 2\sin^2{\omega t\over 2}
-i({\bf n}\cdot{\bf s})\sin\omega t.
\end{eqnarray}
In order to prove this let us note that ${\bf n}\cdot{\bf s}$ is the
operator of the projection of spin-1 on the direction $\bf n$ and it
has three eigenvalue $-1, 0, 1$ with corresponding eigenvectors
$|-1\rangle$, $|0\rangle$, $|1\rangle$ which can play the role of
the basis vectors. An arbitrary vector of state can be written as
linear combination of these vectors. It is enough to
prove formula (\ref{Exp}) only for basis vectors, which are eigenvectors
of ${\bf n}\cdot{\bf s}$ with eigenvalues $-1, 0, 1$. It is easy
to verify that for $\lambda$ which takes only three values $-1, 0,
1$ we have
\begin{eqnarray}\label{elambda}
e^{\lambda x}=
(1-\lambda)(1+\lambda)+{1\over 2}\lambda(\lambda+1)e^x+{1\over 2}\lambda(\lambda-1)e^{-x}.
\end{eqnarray}
Then using (\ref{elambda}) for the unitary operator of evolution
we just obtain (\ref{Exp}).

Let us take the initial vector of state as the eigenvector of $s_z$ with
eigenvalue $-1$
\begin{eqnarray}\label{psii}
|\psi_i\rangle=\left(\begin{array}{c}
                 0 \\
                 0 \\
                 1
               \end{array}\right),
\end{eqnarray}
and reachable final state in form given by (\ref{abc}). Then using
(\ref{psit}), and representation (\ref{Exp}) for the operator of
evolution, and  matrix representation for spin in which $s_z$ is
diagonal, we finally find
\begin{eqnarray}\label{psitabc}
|\psi(t)\rangle=\left(\begin{array}{c}
                  -e^{-i2\phi}\sin^2\theta \sin^2{\omega t\over 2} \\
                  {\sqrt 2}e^{-i\phi}\cos\theta\sin\theta \sin^2{\omega t\over 2}-
                  {i\over\sqrt 2}e^{-i\phi}\sin\theta\sin\omega t \\
                  1-(1+\cos^2\theta)\sin^2{\omega t\over 2}+i \cos\theta\sin\omega t
                \end{array}\right)
\end{eqnarray}
The first component gives the necessary condition that
$|\psi(t)\rangle$ reaches the final state
\begin{eqnarray}\label{eqt}
\sin^2\theta\sin^2{\omega t\over 2}=|a|.
\end{eqnarray}
From (\ref{psitabc}) it follows that the second component depends
on the first one. Substituting $\sin^2{\omega t\over 2}$ from
(\ref{eqt}) into the second component of (\ref{psitabc}) we have
\begin{eqnarray}\label{b}
|b|^2=2|a|(1-|a|).
\end{eqnarray}
Then the normalization condition yields the realtion
\begin{eqnarray}\label{c}
|c|^2=1-|a|^2-|b|^2=(1-|a|)^2.
\end{eqnarray}
Thus we cannot reach arbitrary state, but only such ones which have
components satisfying conditions (\ref{b}) and (\ref{c}). In
addition note that the phases for the second and the third
components are not independent but are related according to
(\ref{psitabc}). If all necessary conditions are
satisfied, then the time of evolution from the initial state to
the allowed final one can be found from (\ref{eqt})
\begin{eqnarray}\label{t}
t_f={4\over\Delta\omega}\arcsin\left({\sqrt{|a|}\over\sin\theta}\right),
\end{eqnarray}
where $\sin\theta>\sqrt{|a|}$, $\hbar \Delta\omega=2\hbar\omega$
is the distance between largest and smallest energy levels.

It is interesting to note that this expression is very similar to
corresponding one for spin-1/2 (see, for instance,
\cite{Bender040}, Eq. (5)). The difference is that (\ref{t}) contains
$\sqrt{|a|}$ instead of $|a|$ and is two times larger (in
\cite{Bender040} $a$ is denoted as $b$ and $\Delta\omega$ is
denoted as $\omega$).

We obtain the minimal time which just corresponds to the quantum
brachistochrone for $\theta=\pi/2$, when magnetic field is
perpendicular to $z$-axis
\begin{eqnarray}
t_{\rm min}={4\over\Delta\omega}\arcsin\left(\sqrt{|a|}\right).
\end{eqnarray}

As an explicit example let us consider the case $|a|=1$, then
$|b|=|c|=0$ and the final state
\begin{eqnarray}\label{psif}
|\psi_f\rangle=\left(\begin{array}{c}
                 1 \\
                 0 \\
                 0
               \end{array}\right)
\end{eqnarray}
is the eigenvector of $s_z$ with the eigenvalue $1$ and is
orthogonal to the initial one (\ref{psii}). In this case we have
the solution for the time only when $\theta=\pi/2$. Thus initial
state (\ref{psii}) evolves to final one (\ref{psif}) only when
magnetic field is perpendicular to $z$-axis. For time of evolution
we have $t_f=t_{\rm min}={2\pi/\Delta\omega}$. This time is two
times longer than the shortest possible time obtained in
\cite{Carlini06,Brody167}. Note that the state vector describing
the evolution in our case is not superposition of initial and
final states. Therefore, it is not strange that the time of
evolution is longer than minimal possible one.

Let us consider the next example with the initial state
\begin{eqnarray}\label{psii2}
|\psi_i\rangle=\left(\begin{array}{c}
                 0 \\
                 1 \\
                 0
               \end{array}\right).
\end{eqnarray}
Now the evolution is given by the state vector
\begin{eqnarray}
|\psi(t)\rangle=\left(\begin{array}{c}
                 -{1\over\sqrt
                 2}e^{-i\phi}\left(2\cos\theta\sin\theta\sin^2{\omega t\over 2}
                 +i\sin\theta\sin\omega t\right) \\
                 1-2\sin^2\theta\sin^2{\omega t\over 2} \\
                 {1\over\sqrt
                 2}e^{i\phi}\left(2\cos\theta\sin\theta\sin^2{\omega t\over 2}
                 -i\sin\theta\sin\omega t\right)
               \end{array}\right)=
\left(\begin{array}{c}
                 -z^* \\
                 1-2\gamma \\
                 z
               \end{array}\right),
\end{eqnarray}
where
\begin{eqnarray}\label{aalpha}
z={1\over\sqrt 2}e^{i\phi}\left(2\cos\theta\sin\theta\sin^2{\omega
t\over 2} -i\sin\theta\sin\omega t\right)=|z|e^{i(\phi-\alpha)},\\
\nonumber |z|^2=2\gamma(1-\gamma), \ \
\gamma=\sin^2\theta\sin^2{\omega t\over 2}, \ \
\tan\alpha={\cos(\omega t/2)\over\cos\theta\sin(\omega t/2)}.
\end{eqnarray}
Finally the evolution of the state vector can be represented
in the form
\begin{eqnarray}\label{Ex2psit}
|\psi(t)\rangle=(1-2\gamma)\left(\begin{array}{c}
                 0 \\
                 1 \\
                 0
               \end{array}\right)+
               \sqrt{2\gamma(1-\gamma)}\left(\begin{array}{c}
                 e^{-i(\phi-\alpha)} \\
                 0 \\
                 e^{i(\phi-\alpha)}
               \end{array}\right),
\end{eqnarray}
where $\gamma$ and $\alpha$ are functions of time as given in
(\ref{aalpha}). Let us consider the final state
\begin{eqnarray}
|\psi_f\rangle={1\over\sqrt2}\left(\begin{array}{c}
                 -1 \\
                 0 \\
                 1
               \end{array}\right)
\end{eqnarray}
which is orthogonal to the initial one (\ref{psii2}). In order to
reach this state we put $\gamma(t_f)=1/2$. This condition gives us
the time of evolution
\begin{eqnarray}
t_f={4\over\Delta\omega}\arcsin\left(1\over\sqrt2\sin\theta\right).
\end{eqnarray}
Then choosing additionally that $\phi=\alpha(t_f)$ we
find that $|\psi(t_f)\rangle=|\psi_f\rangle$.

For $\theta=\pi/2$ we obtain the minimal time of evolution
\begin{eqnarray}
t_{\rm min}={\pi\over \Delta \omega}.
\end{eqnarray}
It is interesting to note that this time is equal to the minimal
possible time which can be obtained according to the statement of
\cite{Carlini06, Brody167} where {\it the state vector of evolution
for minimal possible time belongs to the subspace spanned by the
initial and final state vectors} or, in other words, the vector of
evolution for minimal possible time is a superposition of initial
and final states. In our example, as we see from (\ref{Ex2psit}),
the state vector during evolution does not stay in the subspace
spanned by the initial and final state vectors. Therefore, we can
conclude that {\it in order to achieve the minimal possible time it
is not necessary that during the evolution state vector lies
all the time in the subspace spanned by the initial and final
state vectors}. A pre-constrained family of
Hamiltonians can yield the more complicated evolution, but still
with the optimal time, what can have practical value for experiment.
There is no contradiction with the results
of \cite{Brody167, Brody5587}, the present result is in agreement with 
the fact that to stay on the subspace spanned on the initial and final states one should use the full freedom in the general family of Hamiltonians
for the system under consideration, but what is important, it shows also that staying on the mentioned subspace, is not crucial for implementing
the shortest time evolution.

\end{document}